\documentclass[sigconf,nonacm]{acmart}

\usepackage{graphicx}
\usepackage{textcomp}
\usepackage{xcolor}
\usepackage{subcaption}
\usepackage{svg}
\usepackage{float}
\usepackage{stfloats}
\usepackage{afterpage}
\usepackage{multicol}
\usepackage{array}
\usepackage{mwe}
\usepackage{multirow}
\usepackage{arydshln}
\usepackage[thinc]{esdiff}
\usepackage{booktabs}
\usepackage{pifont}
\usepackage{hyperref}
\usepackage{enumitem}
\usepackage{tablefootnote}
\usepackage{siunitx}
\usepackage{amsmath}
\usepackage[utf8]{inputenc}
\usepackage{makecell}
\usepackage{caption}
\usepackage{tikz}
\usetikzlibrary{arrows.meta, positioning, fit, calc, decorations.pathreplacing, shapes.geometric, backgrounds, patterns}

\definecolor{hwblue}{HTML}{4472C4}
\definecolor{hwgreen}{HTML}{70AD47}
\definecolor{hworange}{HTML}{ED7D31}
\definecolor{hwgray}{HTML}{A5A5A5}
\definecolor{hwlblue}{HTML}{D6E4F0}
\definecolor{hwlgreen}{HTML}{E2EFDA}
\definecolor{hwlorange}{HTML}{FBE5D6}
\definecolor{hwlgray}{HTML}{E8E8E8}
\definecolor{hwpurple}{HTML}{7B68EE}
\definecolor{hwlpurple}{HTML}{E8E0FF}

\AtBeginDocument{}

\begin{document}

\title[EdgeDAE: Acceleration of Diffusion Action Experts for Tiny VLAs on Edge FPGA-GPU Systems]{EdgeDAE: Acceleration of Diffusion Action Experts for Real-Time Physical AI with Tiny VLAs on Edge FPGA-GPU Systems}

\author{Zhiheng Chen}
\affiliation{\department{Nhu Department of Electrical Engineering and Computer Science}\institution{University of California, Irvine}\city{Irvine}\state{CA}\country{USA}}
\email{zhihenc5@uci.edu}

\author{Ye Qiao}
\affiliation{\department{Nhu Department of Electrical Engineering and Computer Science}\institution{University of California, Irvine}\city{Irvine}\state{CA}\country{USA}}
\email{yeq6@uci.edu}

\author{Mohammad Abdullah Al Faruque}
\affiliation{\department{Nhu Department of Electrical Engineering and Computer Science}\institution{University of California, Irvine}\city{Irvine}\state{CA}\country{USA}}
\email{alfaruqu@uci.edu}

\author{Sitao Huang}
\affiliation{\department{Nhu Department of Electrical Engineering and Computer Science}\institution{University of California, Irvine}\city{Irvine}\state{CA}\country{USA}}
\email{sitaoh@uci.edu}

\begin{abstract}
Physical AI models such as Vision-Language-Action (VLA) architectures
enable generalist robotic policies through large-scale transformer backbones and
diffusion-based action decoders. While edge GPU platforms excel at parallelizing
the compute-intensive vision-transformer workloads, they exhibit fundamental
limitations for the Diffusion Action Expert (DAE) module: the iterative denoising
process requires repeated parameter loading from DRAM across multiple steps,
resulting in memory-bound performance where the GPU's massive computational
throughput remains underutilized. This mismatch between DAE's I/O-intensive
characteristics and GPU's compute-centric architecture motivates a heterogeneous
acceleration approach.
This paper presents \textbf{EdgeDAE}, a heterogeneous FPGA-GPU
system that strategically partitions workloads based on computational
characteristics. We offload the perception-heavy vision-transformer to GPU while
accelerating DAE inference on FPGA through complete on-chip parameter storage in
BRAM/URAM. This architecture eliminates the memory bottleneck by co-designing
quantization strategies, fixed-point arithmetic, and hardware-efficient random
number generation for the FPGA fabric. Compared to an edge GPU baseline, EdgeDAE reduces end-to-end inference latency by 52.5\% for
Octo-Small and 38.7\% for Octo-Base, with up to $2.10\times$ higher throughput;
compared to a consumer GPU (RTX~4090), it
achieves ${\sim}17\times$ higher energy efficiency.
\end{abstract}

\keywords{Vision-language-action model, FPGA acceleration, diffusion policy, edge AI, quantization}

\maketitle

\section{Introduction}

Physical AI enables robots to understand and interact with the physical world through embodied intelligence~\cite{yan2025robotic}. Central to this advance are Vision-Language-Action (VLA) models~\cite{brohan2022rt,zitkovich2023rt,team2024octo}, which integrate visual perception, language understanding, and action generation into unified transformer architectures. By pre-training on large-scale robot trajectory datasets such as Open X-Embodiment with over 800K demonstrations across diverse tasks and morphologies~\cite{o2024open}, VLA models achieve remarkable generalization capabilities, enabling zero-shot transfer to new tasks, objects, and environments through natural language instructions or visual goal specifications.

Despite their impressive capabilities, deploying VLA models on edge platforms remains challenging due to their substantial computational and memory requirements. While recent efforts have introduced compact variants---such as the Octo model family achieving orders-of-magnitude size reduction compared to earlier VLA models---real-time inference still demands high-end consumer-grade GPUs. For instance, Octo-Small achieves 17\,it/s and Octo-Base 13\,it/s on an NVIDIA RTX 4090 GPU~\cite{team2024octo}, consuming approximately 200\,W of power~\cite{NVIDIA_RTX4090}. Such power and cost constraints preclude deployment on mobile manipulation platforms, warehouse robots, and field scenarios with strict power, thermal, and cost limitations.

The Octo architecture exposes a workload partitioning opportunity. Its vision-language transformer is compute-intensive and well matched to an edge GPU, while the Diffusion Action Expert (DAE) is a compact MLP-based diffusion policy~\cite{chi2023diffusion,ho2020denoising} executed repeatedly across denoising steps. Despite containing only a small fraction of total model parameters, the DAE becomes a bottleneck on GPUs because iterative sampling reloads parameters from DRAM and leaves wide GPU datapaths underutilized~\cite{chen2023speed,li2023q}. The same compact parameter footprint makes the DAE suitable for FPGA deployment: all weights and diffusion constants can reside in BRAM/URAM, eliminating external memory traffic during the denoising loop. This motivates EdgeDAE's heterogeneous partition: keep perception on GPU and offload the latency-sensitive DAE to FPGA.

This paper presents \textbf{EdgeDAE}, a heterogeneous FPGA-GPU system that accelerates Octo VLA inference on edge platforms through co-designed hardware-software optimizations. We deploy EdgeDAE on a Jetson Orin Nano + Xilinx Kria KR260 heterogeneous platform, where the Jetson handles perception-heavy vision-transformer processing while KR260 accelerates the DAE module. Our key contributions are:

\begin{itemize}[leftmargin=*,nosep]
\item \textbf{Adaptive quantization decoupled from model scale:} We develop per-channel INT8 and NF4-g16 mixed-precision~\cite{dettmers2024qlora} quantization strategies, each deployable across both Octo-Small and Octo-Base to yield four hardware configurations. NF4-g16 halves URAM consumption relative to INT8---from 56 to 28 tiles on Octo-Small---while maintaining or improving action prediction quality. On Octo-Base, NF4-g16 achieves 6.84\% lower mean squared error than the FP32 baseline through implicit regularization.

\item \textbf{Highly-parallelized diffusion sampling architecture:} We implement a parallel Gaussian random number generator pipeline with 4-way Xorshift128 states and Box-Muller transform for 20-step diffusion inference. A unified Quantized Linear Unit kernel with flattened loop structure supports both INT8 direct multiply-accumulate at $P{=}128$ MACs/cycle and NF4 lookup-table-based dequantization at $P{=}64$ MACs/cycle through dual-bank URAM/BRAM access, achieving II=1 pipelined throughput with decoupled SiLU activation.

\item \textbf{End-to-end heterogeneous system integration:} Offloading the I/O-bound DAE to KR260 FPGA at 250\,MHz while retaining compute-bound vision-transformers on Jetson Orin Nano reduces end-to-end latency by 52.5\% for Octo-Small, from 104.18\,ms to 49.50\,ms with $2.10\times$ throughput gain, and by 38.7\% for Octo-Base, from 142.59\,ms to 87.39\,ms with $1.63\times$ throughput gain. EdgeDAE-Small achieves $17.3\times$ higher energy efficiency than an RTX~4090, reaching 1.467 versus 0.085\,it/J at TDP upper bound.
\end{itemize}

\section{Related Work}
\label{sec:related}

Prior diffusion acceleration primarily targets large image-generation
denoisers. GPU-oriented work reduces sampling overhead or applies
post-training quantization, but still assumes a high-bandwidth memory system
for large U-Net or transformer denoisers~\cite{chen2023speed,li2023q}. FPGA
accelerators such as SDA~\cite{10705618} and Diff-Acc~\cite{10.1145/3728470}
improve energy efficiency for Stable Diffusion or unconditional diffusion
models, yet their target networks are orders of magnitude larger than Octo's
DAE and therefore continue to rely on off-chip memory. Embodied-AI hardware
systems such as Dadu-Corki~\cite{huang2025dadu} accelerate broader robot
manipulation pipelines, but do not isolate the compact diffusion action head
as a heterogeneous partitioning target. Beyond diffusion policies, a broader
line of edge-FPGA accelerators targets efficient LLM and transformer
inference~\cite{zhang2025pdswap,10.1145/3748173.3779191,qiao2025cobraalgorithmarchitecturecooptimizedbinary}.

EdgeDAE addresses a different operating point: the DAE is small enough for
complete FPGA on-chip storage, but its repeated 20-step denoising loop is
latency-critical on edge GPUs. This enables a specialized datapath with
integrated Gaussian random number generation, quantized MLP execution, and
precomputed diffusion constants.

\section{Background}
\label{sec:background}

Octo uses a blockwise-causal transformer to process task tokens, temporal
observations, and action readout tokens~\cite{team2024octo}. Language
conditioning is encoded once per episode by T5-Base~\cite{raffel2020exploring};
per-step inference runs image tokenization, the transformer backbone, and a
diffusion action head. Table~\ref{tab:octo_small_specs} summarizes the two
model variants. Both share a three-block MLP ResNet DAE with hidden dimension
$D_h{=}256$, feedforward dimension $D_f{=}1024$, action dimension $D_a{=}28$,
and $T{=}20$ DDPM steps. The main model-scale difference is the DAE input
projection: $D_{\text{in}}{=}444$ for Small and $D_{\text{in}}{=}828$ for Base.

\begin{table}[t]
\centering
\caption{Octo Model Architecture Specifications}
\label{tab:octo_small_specs}
\resizebox{0.84\linewidth}{!}{
\begin{tabular}{l c c c}
\toprule
\textbf{Component} & \textbf{Octo-Small} & \textbf{Octo-Base} & \textbf{Details} \\
\midrule
Observation Encoders & 3.74M & 5.36M & tokenizers, pos.\ emb. \\
BlockTransformer & 21.29M & 85.06M & 12 layers, ViT-S/B dim \\
Diffusion Action Head & 1.70M & 1.80M & 20 diffusion steps \\
Other Components & 0.31M & 0.61M & lang.\ proj., readouts \\
\midrule
\textbf{Total} & \textbf{27M} & \textbf{93M} & excl.\ T5 encoder \\
\bottomrule
\end{tabular}}
\end{table}

\subsection{Diffusion Action Head}

The DAE models action generation as reverse diffusion conditioned on the
transformer readout. At timestep $t$, the MLP predicts noise
$\hat{\boldsymbol{\epsilon}} =
f_\theta(\mathbf{r}_{\text{out}},\mathbf{x}_t,t)$ and updates:
\begin{equation}
\mathbf{x}_{t-1} =
\frac{1}{\sqrt{\alpha_t}}\left(\mathbf{x}_t -
\frac{1-\alpha_t}{\sqrt{1-\bar{\alpha}_t}}\hat{\boldsymbol{\epsilon}}\right)
+ \sqrt{\beta_t}\mathbf{z}_t ,
\label{eq:ddpm_update}
\end{equation}
where $\mathbf{z}_t\sim\mathcal{N}(0,I)$ for $t>0$. Thus each control action
requires 20 MLP forward passes, schedule coefficient lookup, and fresh Gaussian
noise vectors. The MLP consists of an input projection, three residual blocks
with LayerNorm, Linear--SiLU--Linear sublayers, and an output projection to the
28-dimensional action vector.

\subsection{Edge GPU vs.\ Edge FPGA}
\label{sec:gpu_fpga}

This repeated denoising loop is poorly matched to an edge GPU. The Jetson Orin
Nano integrates 1024 CUDA cores, 32 Tensor Cores, 8\,GB LPDDR5, and a 2\,MiB
GPU L2 cache~\cite{NVIDIA_JetsonOrinNanoSuper,NVIDIA_OrinNano_L2}. The DAE
footprint reaches approximately 1.8\,MB at INT8 precision, occupying nearly all
nominal L2 capacity before activations and other working data. Because GPU L2
is shared rather than explicitly pinned, weight residency cannot be assumed
across all 20 steps, and the narrow workload does not saturate the GPU's SIMT
datapath.

The KR260 FPGA has lower raw arithmetic throughput but exposes programmable
on-chip BRAM/URAM storage~\cite{AMD_KR260}. EdgeDAE exploits this by storing
all DAE weights, biases, quantization scales, LayerNorm parameters, time
encodings, and DDPM schedule constants on chip. For the compact Octo DAE, this
turns repeated memory-bound denoising into a deterministic compute pipeline,
while the larger vision-language transformer remains on the GPU.

\section{Architecture Design of EdgeDAE}
\label{sec:architecture}
\subsection{EdgeDAE Design Methodology}

\begin{figure}[t]
    \centering
    \includegraphics[width=\linewidth]{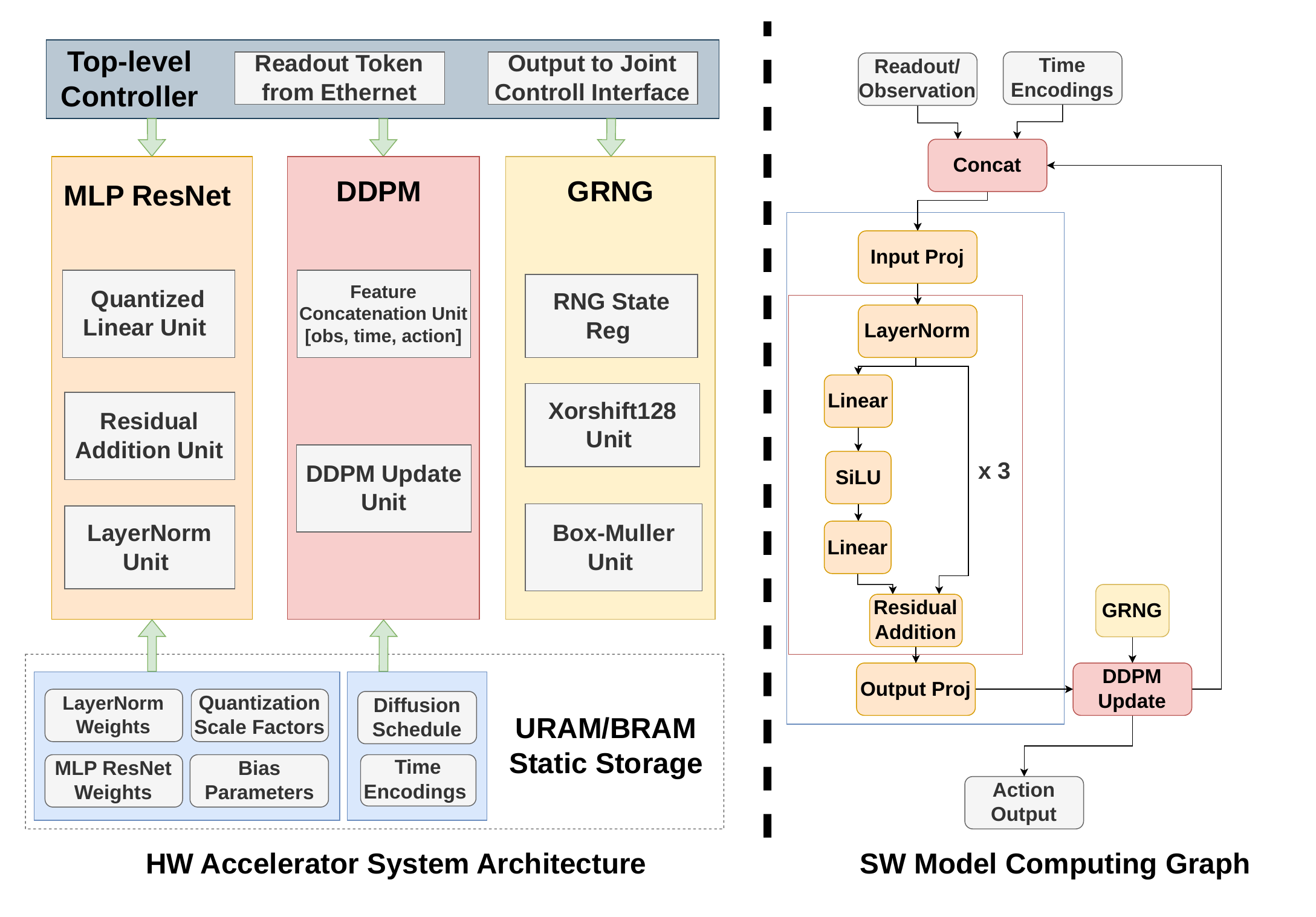}
    \caption{EdgeDAE system architecture (left) and DDPM inference computing graph (right).}
    \Description{Block diagram showing Jetson GPU feature extraction connected to FPGA diffusion action head modules and the DDPM inference dataflow.}
    \label{fig:systemarc}
\end{figure}

As analyzed in Section~\ref{sec:gpu_fpga}, the iterative DDPM denoising process is memory-bound on edge GPUs because its parameter footprint is near the limited L2 cache capacity, leaving little cache space for activations and other working data. EdgeDAE addresses these limitations through a fully self-contained FPGA architecture with complete on-chip parameter storage, integrated random number generation within FPGA logic, and pipelined diffusion execution with pre-loaded constants enabling continuous execution without dynamic memory fetches.

\begin{figure}[t]
    \centering
    \includegraphics[width=\linewidth]{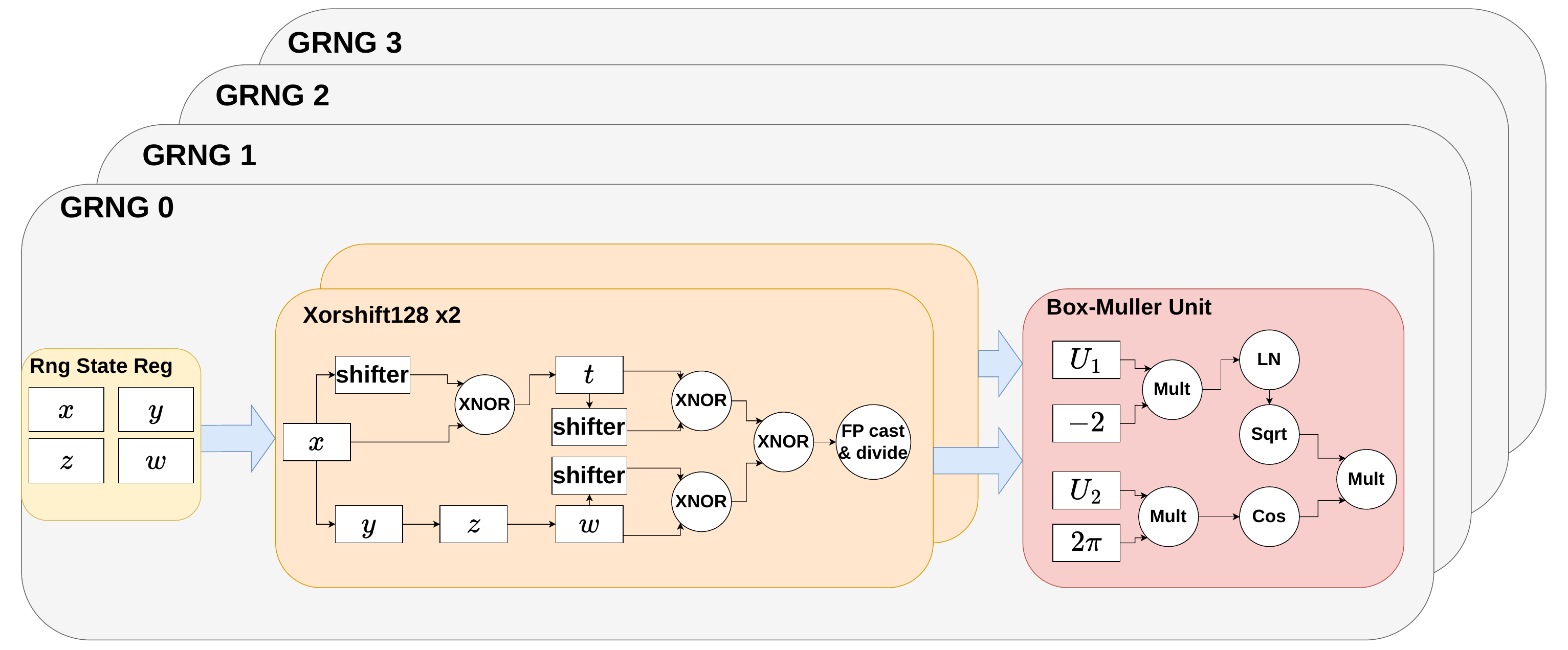}
    \caption{Vectorized GRNG unit.}
    \Description{Hardware block diagram of parallel Xorshift random streams feeding Box-Muller transform lanes for Gaussian noise generation.}
    \label{fig:GRNG}
\end{figure}

The EdgeDAE architecture overview is shown in Fig.~\ref{fig:systemarc}. The left panel depicts the HW accelerator, which comprises three core processing units---the MLP ResNet unit, the DDPM unit, and the Gaussian Random Number Generator (GRNG) unit---managed by a top-level controller that receives readout tokens via Ethernet and outputs joint control commands. All quantized model parameters---weights, biases, quantization scale factors, LayerNorm parameters, time encodings, and diffusion schedule constants---reside permanently in on-chip URAM/BRAM static storage, eliminating external memory dependencies during inference. The right panel illustrates the corresponding software computing graph: observation features and time encodings are concatenated, projected, and processed through three residual MLP blocks, each containing LayerNorm, Linear, SiLU, Linear, and residual addition, followed by an output projection. The GRNG injects fresh Gaussian noise and the DDPM update rule advances the denoising trajectory at each of the $T{=}20$ diffusion steps, producing the final action output.

\subsection{GRNG Unit Design}

The GRNG unit supplies the fresh Gaussian noise vectors required by each DDPM
step without host intervention. It uses four independent Xorshift128 streams,
each with a dedicated 128-bit state, and feeds pairs of uniform samples into a
Box-Muller transform. Complete state partitioning lets all four lanes advance
without contention, producing four Gaussian samples per cycle after pipeline
fill. Since the Octo action vector has $D_a{=}28$ elements, each timestep noise
vector is generated in seven vector blocks and remains off the MLP-dominated
critical path.

\subsection{MLP ResNet Design}

\begin{figure}[t]
    \centering
    \includegraphics[width=\linewidth]{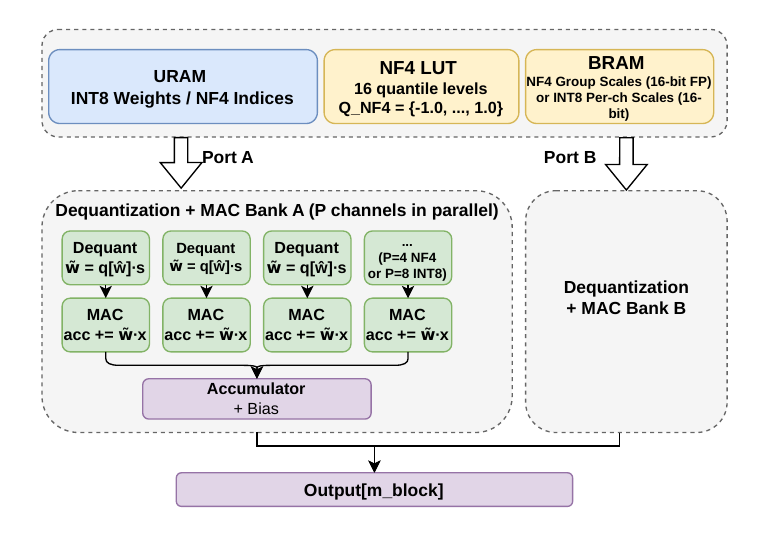}
    \caption{Wide QLU compute architecture.}
    \Description{Datapath diagram for quantized linear unit memory reads, dequantization, parallel MAC arrays, and output accumulation.}
    \label{fig:nf4_hw_arch}
\end{figure}

We denote the input dimension as $D_{\text{in}}$, which equals 444 for Octo-Small and 828 for Octo-Base, along with hidden dimension $D_h{=}256$, feedforward dimension $D_f{=}1024$, action dimension $D_a{=}28$, diffusion timesteps $T{=}20$, and quantization bit-width $B_q$ of 8 for INT8 and 4 for NF4.

The two quantization strategies present a principled latency--resource trade-off. INT8 per-channel stores 8-bit weights directly in URAM, enabling a wide datapath with $P{=}128$ MACs/cycle across 8 output channels $\times$ 16 inputs and the lowest DAE latency (1.38\,ms on Small, 1.80\,ms on Base). NF4-g16, by contrast, packs 4-bit weight indices but requires additional per-group 16-bit scale factors in BRAM, expanding per-channel memory bandwidth and limiting parallelism to $P{=}64$ MACs/cycle across 4 output channels $\times$ 16 inputs---roughly doubling DAE latency (2.87\,ms on Small, 3.14\,ms on Base). In return, NF4-g16 halves URAM consumption from 56 to 28 tiles on Small, freeing on-chip resources for larger models within the fixed FPGA budget.

\subsubsection{Quantization Strategy}

EdgeDAE supports two interchangeable quantization strategies applicable to both Octo-Small and Octo-Base: per-channel INT8 provides a simple, high-throughput datapath, while NF4-g16 mixed-precision achieves lower URAM occupancy at the cost of additional BRAM for group scales.

\textbf{INT8 Per-Channel Quantization.} All diffusion action head weights are quantized to 8-bit integers using per-channel absmax scaling. For each output channel $c$ of weight matrix $\mathbf{W}$:
\begin{equation}
\label{eq:int8}
\left\{
\begin{aligned}
s_c &= \max_{j} |w_{c,j}| \,/\, 127, \\
\hat{w}_{c,j} &= \text{clamp}\!\left(\left\lfloor w_{c,j} / s_c \right\rceil, -128, 127\right), \\
\tilde{w}_{c,j} &= s_c \cdot \hat{w}_{c,j},
\end{aligned}
\right.
\end{equation}
where $\lfloor \cdot \rceil$ denotes rounding to the nearest integer. The per-channel granularity ensures accurate reconstruction without grouping overhead, and the 8-bit integer MAC operations map directly to DSP primitives, enabling high parallelism ($P{=}128$ MACs/cycle) with simple arithmetic.

\textbf{NF4-g16 Mixed-Precision Quantization.} When the parameter footprint warrants more aggressive compression, EdgeDAE employs NormalFloat 4-bit (NF4)~\cite{dettmers2024qlora}, which leverages the empirical observation that pre-trained neural network weights follow approximately normal distributions. The 16 quantization levels are defined as the quantiles of $\mathcal{N}(0,1)$:
\begin{equation}
q_i = \Phi^{-1}\!\left(\frac{2i+1}{32}\right), \quad i = 0, 1, \ldots, 15
\label{eq:nf4_levels}
\end{equation}
where $\Phi^{-1}$ denotes the inverse cumulative distribution function (CDF) of the standard normal distribution. This formulation provides information-theoretically optimal 4-bit representation for normally-distributed data~\cite{dettmers2024qlora}.

For each weight matrix $\mathbf{W}$, weights are partitioned into contiguous groups of size $g{=}16$ along the input dimension. Within each group $\mathcal{G}_k$, the quantization and dequantization procedure is:
\begin{equation}
\label{eq:nf4}
\left\{
\begin{aligned}
s_k &= \max_{j \in \mathcal{G}_k} |w_j|, \\
\hat{w}_j &= \arg\min_{q_i \in \mathcal{Q}_{\text{NF4}}} \left| \frac{w_j}{s_k} - q_i \right|, \\
\tilde{w}_j &= q_{\hat{w}_j} \cdot s_k,
\end{aligned}
\right.
\end{equation}
where $\mathcal{Q}_{\text{NF4}} = \{-1.0, -0.696, -0.525, \ldots, 0.723, 1.0\}$ denotes the 16 NF4 quantile levels, $s_k$ is the per-group absmax scale, and $\hat{w}_j$ is the 4-bit quantized index stored as a nibble.

\textbf{Mixed-Precision Layer Assignment.} Under NF4-g16, the I/O projection layers, specifically init\_proj at $D_{\text{in}}{\times}D_h$ and final\_dense at $D_h{\times}D_a$, retain per-channel INT8 quantization (Eq.~\ref{eq:int8}), as they account for only ${\sim}5\%$ of parameters. The three ResNet blocks' linear layers ($D_h{\times}D_f$ and $D_f{\times}D_h$) constitute ${\sim}95\%$ of parameters with well-conditioned normal distributions ideal for NF4-g16. This mixed-precision assignment is identical for both Octo-Small and Octo-Base. 

\subsubsection{Memory Architecture and Loop Flatten Optimization}

The MLP ResNet architecture consists of an initial projection from $D_{\text{in}}$ to $D_h$, three residual blocks mapping $D_h$ to $D_f$ and back to $D_h$, and an output projection from $D_h$ to $D_a$. All weights are stored in on-chip URAM with dual-port access, while biases and scale factors reside in BRAM, both using offset-based addressing to eliminate data movement between layers. The unified memory layout follows $[\text{weights}_\text{all}] = [\mathbf{W}_\text{init},\; \mathbf{W}_{\text{block}_0}^{(1)},$$\mathbf{W}_{\text{block}_0}^{(2)},\; \ldots,\; \mathbf{W}_\text{out}]$, with analogous layouts for $[\text{bias}_\text{all}]$ and $[\text{scale}_\text{all}]$.

The matrix multiplication kernel uses a manually flattened loop structure to
keep initiation interval II=1 across the different matrix shapes in the input
projection, residual blocks, and output projection. QLU maps the two-dimensional
output-block/input-block iteration space to one pipelined iterator, caches input
blocks, initializes accumulators with bias at the start of each output block,
and applies a shared dequantization abstraction. For INT8,
$\textsc{Dequant}(w,s)=s_c\hat{w}$ uses direct per-channel scaling; for
NF4-g16, $\textsc{Dequant}(w,s)=\mathcal{Q}_{\text{NF4}}[\hat{w}]s_k$ uses a
16-entry lookup table and per-group scale.

Three key hardware optimizations further improve QLU throughput.
\textbf{(1) INT8 direct MAC.} Per-channel INT8 requires no LUT: the dequantized value $s_c \cdot \hat{w}$ is computed by a single DSP multiply, and the wide datapath processes $P{=}128$ MACs/cycle across 8 output channels simultaneously, computing 16 input elements $\times$ 8 output channels per cycle. The weight URAM stores 8-bit values directly, yielding $2\times$ memory bandwidth compared to NF4.
\textbf{(2) NF4 LUT pre-scaling.} The per-group scale multiplication is amortized by pre-computing 16 scaled NF4 levels $\hat{q}_i = q_i \cdot s_k$ before the inner MAC loop, reducing each dequantization to a single table lookup without additional multiplications. The wide NF4 datapath processes $P{=}64$ MACs/cycle across 4 output channels, computing 16 input elements $\times$ 4 output channels per cycle, as each 4-bit weight with its group scale occupies more BRAM bandwidth per channel than INT8.
\textbf{(3) Dual-port URAM and wide datapath.} The weight URAM is configured as dual-port via \texttt{ram\_2p} with even- and odd-indexed output blocks assigned to ports A and B respectively, enabling two simultaneous weight reads per cycle. The recurrence distance of $\text{BLOCK}_\text{out}$ cycles exceeds the MAC pipeline latency, guaranteeing II=1 without accumulator hazards. Both INT8 and NF4 datapaths share this dual-port structure; the parallelism difference of $P{=}128$ for INT8 versus $P{=}64$ for NF4 arises from the per-element memory footprint, not from model choice. Figure~\ref{fig:nf4_hw_arch} details the wide parallel datapath of the QLU compute unit.

The SiLU activation $f(x)=x\sigma(x)$ runs in a separate II=1 post-processing
loop using a 256-entry sigmoid lookup table over $[-6,6]$ with spacing 0.046875,
providing a single-cycle approximation for the nonlinear stage. Decoupling
activation from the MAC loop prevents LUT read-port contention with weight and
scale accesses; embedding SiLU lookups in the same pipeline body would create
resource conflicts that degrade initiation interval. The separate activation
loop adds at most $\lceil D_h/\text{BLOCK}_{\text{out}}\rceil$ cycles per layer,
which is negligible relative to the MAC-dominated computation.

\subsubsection{ResNet Block Implementation}

Each residual block implements the structure $\mathbf{x}' = \mathbf{x} + \text{MLP}(\text{LayerNorm}(\mathbf{x}))$, where MLP consists of two linear layers with SiLU activation. The LayerNorm computation normalizes each $D_h$-dimensional vector via $\hat{\mathbf{x}}_j = \gamma_j \frac{\mathbf{x}_j - \mu}{\sqrt{\sigma^2 + \epsilon}} + \beta_j$, with statistics $\mu = \frac{1}{D_h}\sum_j \mathbf{x}_j$ and $\sigma^2 = \frac{1}{D_h}\sum_j (\mathbf{x}_j - \mu)^2$ computed through parallel reduction trees. The residual skip connection bypasses the two-layer MLP computation, with element-wise addition performed on $\text{BLOCK}_\text{out}$-element blocks in parallel, achieving II=1 throughput for the $D_h$-dimensional vector.

\subsection{DDPM Denoising Unit}

The DDPM denoising unit implements the reverse diffusion process across
$T{=}20$ sequential timesteps, transforming Gaussian noise into structured action
predictions. At each timestep $t$, the unit executes
$\mathbf{x}_{t-1}=\frac{1}{\sqrt{\alpha_t}}(\mathbf{x}_t -
\frac{1-\alpha_t}{\sqrt{1-\bar{\alpha}_t}}\boldsymbol{\epsilon}_\theta)+
\sqrt{\beta_t}\mathbf{z}_t$, where $\boldsymbol{\epsilon}_\theta$ is the MLP
ResNet noise prediction and $\mathbf{z}_t\sim\mathcal{N}(\mathbf{0},\mathbf{I})$
is fresh Gaussian noise.

The architectural innovation exploits the deterministic nature of diffusion
schedule parameters and time encodings. The diffusion schedule
$\{\beta_t,\alpha_t,\bar{\alpha}_t\}_{t=0}^{T-1}$ follows a fixed cosine
schedule that remains constant across all inferences, enabling compile-time
computation and storage in on-chip RAM. Time encoding vectors exhibit similar
determinism: for each timestep $t$, the Fourier features
$\mathbf{f}_t=[\sin(2^i t),\cos(2^i t)]_{i=0}^{15}$ and subsequent conditional
encoder transformation produce identical outputs independent of input
observations. Both components depend solely on the timestep index, not on the
current action state or observation embedding, allowing complete pre-computation
and on-chip storage of all time-dependent constants.

The denoising unit maintains all schedule parameters and pre-computed time encodings in on-chip storage, enabling direct lookup operations during the iterative process. At each timestep, the unit retrieves the corresponding time encoding vector, concatenates it with the observation embedding and current action to form the network input, and applies the DDPM update rule using the pre-loaded schedule coefficients $\{\sqrt{\alpha_t}, \frac{1-\alpha_t}{\sqrt{1-\bar{\alpha}_t}}, \sqrt{\beta_t}\}_{t=0}^{T-1}$, which are pre-computed from the base schedule parameters. 

The update hardware architecture implements the denoising formula through three parallel computation stages operating on $D_a$-dimensional action vectors partitioned into blocks. Stage 1 computes the noise removal term $\frac{1-\alpha_t}{\sqrt{1-\bar{\alpha}_t}} \boldsymbol{\epsilon}_\theta$ via element-wise multiplication with the pre-loaded coefficient, achieving II=1 throughput through vectorized operations. Stage 2 performs action subtraction $\mathbf{x}_t - \frac{1-\alpha_t}{\sqrt{1-\bar{\alpha}_t}} \boldsymbol{\epsilon}_\theta$ and scales by $\frac{1}{\sqrt{\alpha_t}}$ in a single fused operation to minimize intermediate storage. Stage 3 adds the stochastic noise term $\sqrt{\beta_t} \mathbf{z}_t$ when $t > 0$, with the coefficient multiplication fused into the addition. The action clipping operation is integrated into Stage 3 output, applying $\text{clip}(\mathbf{x}_{t-1}, -5, 5)$ to maintain valid action ranges without additional passes. This three-stage pipeline maintains deterministic latency through fully pipelined arithmetic units with pre-loaded coefficients residing in local registers, ensuring zero external memory dependency throughout the $T$-step iterative process.

\section{Experimental Results and Analysis}
\label{sec:experiments}

We evaluate EdgeDAE along three axes: FPGA resource utilization, end-to-end latency and energy efficiency versus GPU baselines, and action prediction quality under different quantization strategies.

The EdgeDAE diffusion action head is implemented on the KR260 FPGA platform at 250 MHz clock frequency. The complete Octo VLA system employs a heterogeneous architecture with Jetson Orin Nano in MAXN\_SUPER mode, executing the vision transformer encoder using PyTorch~2.3.0 with \texttt{torch.compile} optimization, while the KR260 accelerates the diffusion action head with quantized weights stored entirely on-chip. The non-diffusion GPU pipeline---language encoder, image encoder, and vision-language transformer---is compiled as a single \texttt{torch.compile} unit with \texttt{reduce-overhead} mode, enabling cross-module optimization. Both Octo-Small with ViT-S 384-dim backbone and Octo-Base with ViT-B 768-dim backbone are profiled using CUDA event timing over 200 iterations after 50 warmup iterations. The two platforms communicate via 1~Gbps Ethernet using UDP protocol to transfer the readout token---$1 \times 384$ for Octo-Small and $1 \times 768$ for Octo-Base---with measured latency of 70\,$\mu$s. Baseline GPU performance metrics for RTX 4090 are obtained from the official Octo GitHub repository\cite{team2024octo}. For the RTX~4090 power consumption, we reference the official NVIDIA product specifications~\cite{NVIDIA_RTX4090}, which rate the maximum Thermal Design Power at 450\,W; we adopt 200\,W as TDP upper bound for the energy efficiency comparison. Table~\ref{tab:latency_comparison} presents end-to-end comparisons for both model variants, while Table~\ref{tab:octo_base_fpga} details the FPGA DAE performance under NF4 and INT8 quantization strategies.
\subsection{Resource Breakdown}
Table~\ref{tab:resource_all} presents post-place-and-route FPGA resource utilization for all four EdgeDAE configurations on the KR260 platform at 250\,MHz. The MLP ResNet module dominates DSP consumption at ${\sim}75\%$ across all configurations. The key resource trade-off between quantization strategies is apparent: INT8 consumes 56 URAM tiles at 87.5\% utilization for direct weight storage, while NF4-g16 halves URAM to 28--32 tiles at 43.8--50.0\% by moving per-group scales to 16-bit fixed-point BRAM, which correspondingly rises to 96.5--97.2\%. NF4 builds also exhibit ${\sim}10\%$ higher LUT utilization due to dequantization logic. All configurations maintain timing closure at 250\,MHz with DSP utilization under 29\%. Table~\ref{tab:octo_base_fpga} summarizes the resulting performance metrics across all four configurations.

\begin{table}[t]
\centering
\caption{FPGA Resource Utilization (Post-P\&R, 250\,MHz) for All EdgeDAE Configurations}
\label{tab:resource_all}
\resizebox{\linewidth}{!}{
\renewcommand{\arraystretch}{0.85}
\begin{tabular}{l cc cc c}
\toprule
& \multicolumn{2}{c}{\textbf{Octo-Small}} & \multicolumn{2}{c}{\textbf{Octo-Base}} & \\
\cmidrule(lr){2-3} \cmidrule(lr){4-5}
\textbf{Resource} & \textbf{INT8} & \textbf{NF4-g16} & \textbf{INT8} & \textbf{NF4-g16} & \textbf{KR260} \\
\midrule
LUT & 46{,}932 (40.1\%) & 58{,}527 (50.0\%) & 46{,}933 (40.1\%) & 58{,}776 (50.2\%) & 117{,}120 \\
FF & 50{,}006 (21.4\%) & 67{,}024 (28.6\%) & 49{,}884 (21.3\%) & 68{,}808 (29.4\%) & 234{,}240 \\
BRAM & 33.5 (23.3\%) & 139 (96.5\%) & 35.5 (24.7\%) & 140 (97.2\%) & 144 \\
URAM & 56 (87.5\%) & 28 (43.8\%) & 56 (87.5\%) & 32 (50.0\%) & 64 \\
DSP & 356 (28.5\%) & 349 (28.0\%) & 356 (28.5\%) & 351 (28.1\%) & 1{,}248 \\
Power (W) & 4.51 & 5.37 & 4.72 & 5.28 & --- \\
\bottomrule
\multicolumn{6}{l}{\footnotesize All totals: Vivado post-place-and-route at 250\,MHz.} \\
\end{tabular}
}
\end{table}

\begin{table}[t]
\centering
\caption{EdgeDAE FPGA Performance: All Four Quantization Configurations at 250\,MHz}
\label{tab:octo_base_fpga}
\resizebox{\linewidth}{!}{
\renewcommand{\arraystretch}{0.85}
\begin{tabular}{l c c c c}
\toprule
\textbf{Metric} & \textbf{\makecell{Small\\INT8}} & \textbf{\makecell{Small\\NF4-g16}} & \textbf{\makecell{Base\\INT8}} & \textbf{\makecell{Base\\NF4-g16}} \\
\midrule
Input Dim & 444 & 444 & 828 & 828 \\
DAE Latency (ms) & 1.38 & 2.87 & 1.80 & 3.14 \\
Quantization & INT8 per-ch & NF4-g16 & INT8 per-ch & NF4-g16 \\
On-chip (MB) & 1.21 & 0.62 & 1.80 & 0.84 \\
URAM Usage & 56 (87.5\%) & 28 (43.8\%) & 56 (87.5\%) & 32 (50.0\%) \\
DSP & 356 (28.5\%) & 349 (28.0\%) & 356 (28.5\%) & 351 (28.1\%) \\
LUT & 46{,}932 (40.1\%) & 58{,}527 (50.0\%) & 46{,}933 (40.1\%) & 58{,}776 (50.2\%) \\
Success Rate & 83.74\% & 82.59\% & 87.51\% & 85.90\% \\
\bottomrule
\end{tabular}}
\end{table}

\begin{figure}[t]
    \centering
    \begin{subfigure}[b]{0.46\linewidth}
        \centering
        \includegraphics[width=\linewidth]{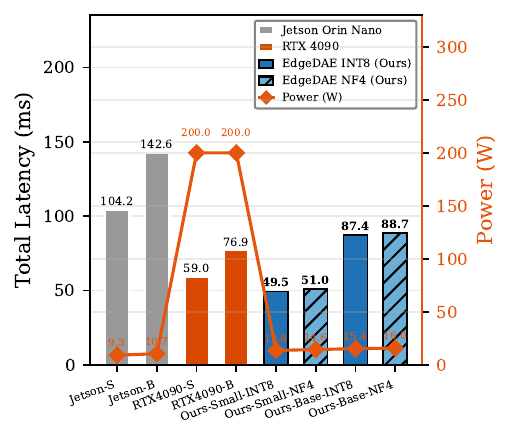}
        \caption{Latency and power comparison.}
        \label{fig:latency_power}
    \end{subfigure}
    \hfill
    \begin{subfigure}[b]{0.46\linewidth}
        \centering
        \includegraphics[width=\linewidth]{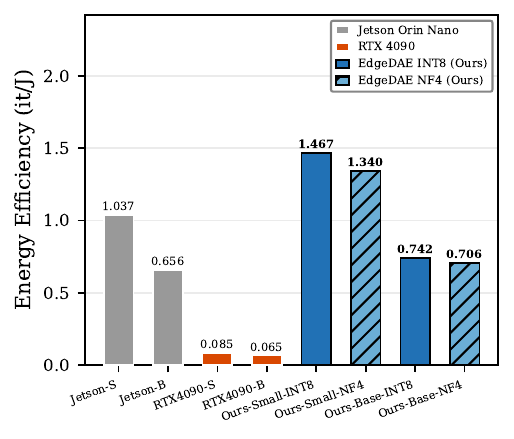}
        \caption{Energy efficiency comparison.}
        \label{fig:energy_efficiency}
    \end{subfigure}
    \caption{Performance comparison across configurations.}
    \Description{Two bar charts comparing latency and power as well as energy efficiency for Jetson, RTX 4090, and EdgeDAE configurations.}
    \label{fig:performance_comparison}
\end{figure}

\begin{figure*}[!t]
    \centering
    \begin{subfigure}[b]{\linewidth}
        \centering
        \includegraphics[width=0.83\linewidth]{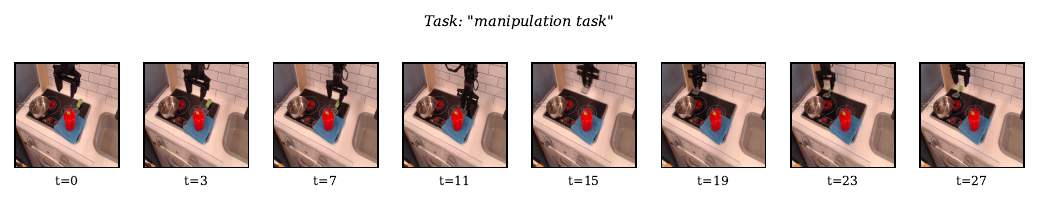}
        \caption{Evaluation episode camera frames.}
        \label{fig:dataset_examples}
    \end{subfigure}
    \\[-2pt]
    \begin{subfigure}[b]{\linewidth}
        \centering
        \includegraphics[width=0.747\linewidth]{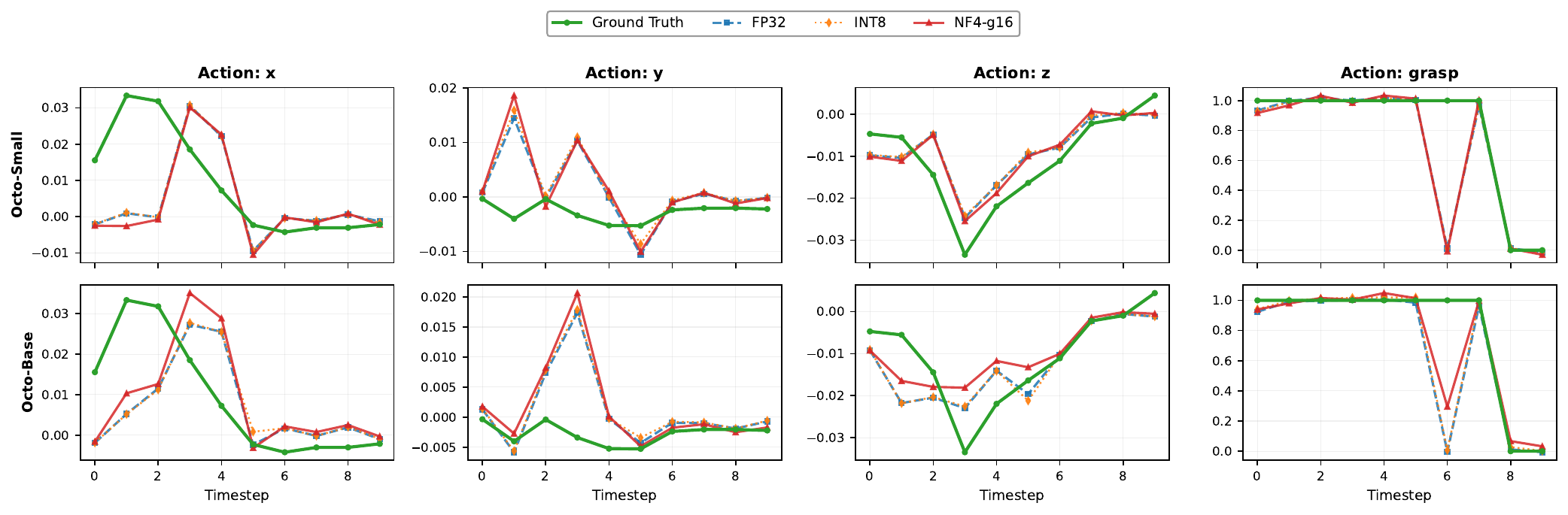}
        \caption{Action trajectory comparison across quantization configurations.}
        \label{fig:action_comparison}
    \end{subfigure}
    \caption{Bridge dataset evaluation and trajectory comparison.}
    \Description{Camera-frame examples from Bridge evaluation episodes and trajectory plots comparing FP32, INT8, NF4, and INT4 action predictions.}
    \label{fig:dataset_and_trajectory}
\end{figure*}

INT8 configurations achieve lower LUT utilization at 40.1\% and FF at 21.3--21.4\% compared to NF4 at 50.0--50.2\% LUT and 28.6--29.4\% FF, due to simpler arithmetic without dequantization overhead. The dual-strategy design is validated: INT8 minimizes latency when URAM budget permits, while NF4-g16 enables deployment of larger models by halving URAM requirements.

\subsection{Latency and Energy Efficiency Comparison}

Table~\ref{tab:latency_comparison} compares end-to-end latency and energy efficiency across the Jetson Orin Nano baseline, RTX~4090~\cite{team2024octo}, and EdgeDAE. The GPU Backbone column reports the combined latency of the language encoder, image encoder, and vision-language transformer compiled as a single \texttt{torch.compile} unit. DAE latency includes FPGA inference plus 70\,$\mu$s UDP communication overhead.
For Octo-Small, EdgeDAE-INT8 reduces DAE latency from 56.06\,ms on GPU to 1.38\,ms on FPGA---a $40.6\times$ speedup---cutting total inference from 104.18\,ms to 49.50\,ms, a 52.5\% reduction that yields 20.20\,it/s throughput, surpassing the RTX~4090 at 17\,it/s. For Octo-Base, EdgeDAE-INT8 accelerates the DAE by $31.7\times$ from 57.00\,ms to 1.80\,ms, reducing total inference from 142.59\,ms to 87.39\,ms---a 38.7\% reduction with $1.63\times$ throughput gain. At this operating point, the ViT-B backbone at 85.59\,ms dominates 97.9\% of EdgeDAE total latency.
In energy efficiency, EdgeDAE-Small-INT8 achieves 1.467\,it/J, which is $17.3\times$ higher than the RTX~4090 at 0.085\,it/J under TDP upper bound. EdgeDAE-Base-INT8 reaches 0.742\,it/J, a $1.13\times$ improvement over the Jetson baseline at 0.656\,it/J. These results confirm that FPGA offloading improves energy efficiency across both model scales, with more pronounced gains when the accelerated module constitutes a larger fraction of total latency.

\begin{table}[t]
\centering
\caption{End-to-End Latency and Energy Efficiency Comparison (Jetson Orin Nano MAXN\_SUPER)}
\label{tab:latency_comparison}
\resizebox{\linewidth}{!}{
\renewcommand{\arraystretch}{0.85}
\begin{tabular}{l c c c c c c}
\toprule
\textbf{Setup} & \textbf{\makecell{GPU \\ Backbone}} & \textbf{\makecell{Diffusion \\ Action \\ Head}} & \textbf{\makecell{Total \\ (ms)}} & \textbf{\makecell{Throughput \\ (it/s)}} & \textbf{\makecell{Power \\ (W)}} & \textbf{\makecell{Efficiency \\ (it/J)}} \\
\midrule
Jetson (Octo-Small) & 48.12 & 56.06 & 104.18 & 9.60 & 9.26 & 1.037 \\
Jetson (Octo-Base) & 85.59 & 57.00 & 142.59 & 7.01 & 10.69 & 0.656 \\
RTX 4090 (Small)~\cite{team2024octo} & - & - & 59.0 & 17.0 & 200.0 & 0.085 \\
RTX 4090 (Base)~\cite{team2024octo} & - & - & 76.9 & 13.0 & 200.0 & 0.065 \\
\textbf{EdgeDAE-Small-INT8 (Ours)} & 48.12 & \textbf{1.38} & \textbf{49.50} & \textbf{20.20} & 13.77 & \textbf{1.467} \\
\textbf{EdgeDAE-Small-NF4 (Ours)} & 48.12 & 2.87 & 50.99 & 19.61 & 14.63 & 1.340 \\
\textbf{EdgeDAE-Base-INT8 (Ours)} & 85.59 & 1.80 & 87.39 & 11.44 & 15.41 & 0.742 \\
\textbf{EdgeDAE-Base-NF4 (Ours)} & 85.59 & 3.14 & 88.73 & 11.27 & 15.97 & 0.706 \\
\bottomrule
\end{tabular}
}
\end{table}

\subsection{Action Quality Comparison}

Both quantization strategies are evaluated on the Bridge dataset test split comprising 1000 episodes and 33{,}188 timesteps. Table~\ref{tab:quantization_ablation} reports ground-truth MSE and success rate with threshold $\tau{=}0.15$ for all configurations.

For Octo-Base, NF4-g16 achieves 6.84\% lower MSE than the FP32 baseline, consistent with implicit regularization effects of quantization noise in diffusion inference~\cite{li2023q}. INT8 preserves near-lossless quality with $-$3.04\% MSE change and 87.51\% success rate. The grouped INT4-g16 baseline drops to 81.70\% success rate, confirming NF4's superiority over uniform 4-bit grids. For Octo-Small, INT8 at $+$0.75\% MSE with 83.74\% success rate and NF4-g16 at $+$2.18\% MSE with 82.59\% success rate both remain within 1.5 percentage points of the FP32 baseline at 84.06\%.

\begin{table}[t]
\centering
\caption{Quantization Accuracy Ablation on Bridge Dataset (1000 episodes, 33{,}188 timesteps)}
\label{tab:quantization_ablation}
\resizebox{\linewidth}{!}{
\renewcommand{\arraystretch}{0.85}
\begin{tabular}{l c c c c}
\toprule
\textbf{Strategy} & \textbf{Bits} & \textbf{GT-MSE} & \textbf{$\Delta$(\%)} & \textbf{Success Rate} \\
\midrule
\multicolumn{5}{c}{\textit{Octo-Base}} \\
\midrule
FP32 baseline & 32 & 0.010203 & --- & 87.49\% \\
\textbf{INT8 per-channel (Ours)} & \textbf{8} & \textbf{0.009893} & $\mathbf{-}$\textbf{3.04} & \textbf{87.51\%} \\
gINT4-g16 mixed & 4 & 0.010695 & $+$4.82 & 81.70\% \\
\textbf{NF4-g16 mixed (Ours)} & \textbf{4} & \textbf{0.009505} & $\mathbf{-}$\textbf{6.84} & \textbf{85.90\%} \\
\midrule
\multicolumn{5}{c}{\textit{Octo-Small}} \\
\midrule
FP32 baseline & 32 & 0.01479 & --- & 84.06\% \\
\textbf{INT8 per-channel (Ours)} & \textbf{8} & \textbf{0.01490} & $+$\textbf{0.75} & \textbf{83.74\%} \\
\textbf{NF4-g16 mixed (Ours)} & \textbf{4} & \textbf{0.01511} & $+$\textbf{2.18} & \textbf{82.59\%} \\
INT4 mixed & 4 & 0.2903 & $+$1863.0 & $\approx$0\% \\
\bottomrule
\end{tabular}}
\end{table}

Fig.~\ref{fig:action_comparison} shows that quantized trajectories closely track FP32 baselines across all configurations, with grasp timing preserved at high fidelity.

\section{Conclusion}
\label{sec:conclusion}
EdgeDAE offloads the diffusion action head to FPGA while retaining vision-language processing on an edge GPU. INT8/NF4-g16 quantization keeps all weights on chip; NF4-g16 halves URAM usage relative to INT8 while preserving action quality. Against the Jetson-only baseline, EdgeDAE cuts end-to-end latency by 52.5\% for Octo-Small and 38.7\% for Octo-Base, while delivering $17.3\times$ higher energy efficiency than an RTX~4090.

\clearpage
\bibliographystyle{ACM-Reference-Format}
\bibliography{references}

\end{document}